\documentclass[conference,compsoc]{IEEEtran}

\usepackage[T1]{fontenc}
\usepackage{microtype}
\usepackage{amsmath,amssymb,amsthm,mathtools}
\usepackage{enumitem}
\usepackage{hyperref}
\usepackage{cleveref}
\usepackage{xspace}
\usepackage{wrapfig}
\usepackage{tabularx}
\usepackage{booktabs}
\usepackage{array}
\usepackage{float}

\usepackage{graphicx}
\usepackage[caption=false,font=normalsize,
            labelfont=sf,textfont=sf]{subfig}

\usepackage{tikz}
\usetikzlibrary{positioning, fit, arrows, calc}
\usetikzlibrary{arrows.meta}

\usepackage{adjustbox}

\usepackage{nicefrac}
\usepackage{booktabs}

\usepackage{colortbl}

\newtheorem{proposition}{Proposition}
\newtheorem{theorem}[proposition]{Theorem}

\theoremstyle{remark}

\newtheorem{definition}{Definition}

\newcommand{\IMR}{QUASI\xspace}
\newcommand{\QUASI}{\IMR}
\newcommand{\unforge}{Weak-$f$-Unforgeability\xspace}

\newcommand{\dmin}{\ensuremath{D_{min}}\xspace}

\hypersetup{
  colorlinks=true,
  linkcolor=blue!50!black,
  urlcolor=blue!50!black,
  pdftitle={QUASI: Post-Quantum Quorum Attestation of Pre-Validated Statements}
}

\title{QUASI: Post-Quantum Quorum Attestation of Pre-Validated Statements}

\author{
\IEEEauthorblockN{Maksym Ioffe}
\IEEEauthorblockA{
University of California, Santa Barbara\\
maksym@ucsb.edu
}
\and
\IEEEauthorblockN{Dahlia Malkhi}
\IEEEauthorblockA{
University of California, Santa Barbara\\
dahliamalkhi@gmail.com
}
}

\begin{document}
\maketitle

\begin{abstract}
 We present QUASI, a newly conceived class of hash-based, post-quantum-secure protocols that enable a group of parties to produce a publicly verifiable attestation about a pre-validated statement. QUASI produces attestations that reduce the size and verification cost to a fraction of a full hash-based signatures collection, without relying on SNARKs or a trusted setup.

QUASI rests on a surprising discovery: each party need not sign the entire statement. Authenticating only a small subset of symbols suffices for a quorum to attest to the whole. We turn this insight into a randomized construction that achieves reductions of more than 10x in per-party contribution size and hash computations for generation and verification, compared with full hash-based signatures.

QUASI is a generic primitive supporting diverse applications. Operationally, QUASI requires no additional interactive setup or preprocessing
beyond independent key generation by parties for the underlying hash-based authentication
scheme.
Two principal use cases are external attestation for signature-free consensus, enabling clients to verify statements about committed state; and preprocessing for SNARK-based aggregation, compounding QUASI’s savings with succinct-proof compression. 
\end{abstract}

\section{Introduction}

The need for aggregating post-quantum (PQ) signatures has been articulated in several
reports~\cite{BonehGentryLynnShacham03,coinbase-quantum2026}.
NIST's recent Threshold Call identifies aggregation into
sublinear-size results, quantum resistance, and ZKP-friendliness
as features of interest~\cite[Sections 3.2 and 11.1]{NIST8214C}. Hash-based signatures, such as Lamport-OTS~\cite{LOTS} and WOTS~\cite{WOTS}, are central to post-quantum migration~\cite{aaronson2004collision}. 
Such primitives also underpin hash-based signature management
schemes such as XMSS~\cite{xmss} and SLH-DSA~\cite{slhdsa},
which play a prominent role in PQ migration.
However, hash-based signature schemes lack BLS-style native aggregation:
the prevailing approach uses SNARK-based aggregation to compress
full signatures contributed by every party~\cite{dkkw}.
Thus, aggregating 100 hash-based signatures today means either
concatenating them ($100\times$ the size and $100\times$ the
verification work) or proving all 100 verifications inside a
hash-based SNARK, as in Ethereum's plan~\cite{ethereum}.
Recent constructions extend hash-based signatures to threshold signing, but require trusted-dealer preprocessing and share reconstruction~\cite{kelsey2025threshold,BonehEtAl2026PRAWNS}.

We present \QUASI{}, 
a newly conceived class of hash-based attestation protocols. 
QUASI rests on a surprising discovery: 
a group of parties can produce a compact, publicly verifiable, PQ-secure attestation without each party contributing a full signature. In particular,
each party need not sign the entire statement.
The \QUASI{} family is based purely on hashing and incorporates
error correction. Its protocols have small per-party contributions
compared with full hash-based signatures, have low
quorum-attestation generation and verification costs, and do not
rely on SNARKs for aggregation.

%

\paragraph{\textbf{Core insight and terminology}}
\QUASI{} is not a signature aggregation scheme, nor
does it provide the guarantees of threshold signatures.
It addresses a weaker task: honest parties in a quorum-based
system have already validated and agreed on a single statement
per context, and the goal is to attest to external parties that
this statement is valid.

This distinction enables a previously unexplored approach: each party
authenticates only a small subset of the symbols that make up the statement, rather than
providing a full signature. This is possible because hash-based
authentication treats individual input symbols independently.
We call a party's contribution a \textit{quasi-signature share} (qShare); a quorum of qShares produces a publicly
verifiable \textit{joint quorum attestation} (qAttest).
Together, these contributions suffice to verify the entire
pre-validated statement validity.
The number of symbols that a qShare needs to authenticate
decreases as the number of parties increases, reaching as few
as one symbols per party.

\paragraph{\textbf{Use cases}}

\QUASI is a generic primitive supporting diverse applications.
The need to make protocol outcomes verifiable by external
parties arises across a broad range of quorum-based systems:
low-latency payment and transaction certification, as in
FastPay, Sui, and Stingray~\cite{fastpay,sui-lutris,stingray};
reliable and atomic broadcast, including Bracha's protocol,
the protocols of Cachin et al., and
Rampart~\cite{bracha1987,cachin2001broadcast,rampart};
and Byzantine consensus, from the foundational Byzantine
Generals Problem of Lamport, Shostak, and
Pease~\cite{byzantine-generals}, through the seminal 
PBFT protocol~\cite{pbft}, 
to modern protocols such as
HotStuff~\cite{hotstuff}.
QUASI offers an external attestation layer for these settings
when honest parties have pre-validated the statement being
attested.

Of particular note, \QUASI{} may provide external
attestation for signature-free Byzantine consensus protocols
and blockchains.
These protocols, also known as information-theoretic (IT)
consensus protocols, achieve agreement using authenticated
point-to-point channels. Examples include IT-HotStuff,
Forget-IT, TetraBFT, Simple-IT, Quintus, Quantum
Cadence~\cite{IT-hotsuff,forgetit,tetrabft,simpleit,quintus,quantumcadence}, and others.
Although these protocols provide post-quantum-secure agreement,
their internal authentication does not produce publicly
verifiable attestations about committed state.
\QUASI{} fills this gap by providing an external attestation
layer, enabling clients to verify statements about committed state.

Another use case compounds the \QUASI{} advantage with SNARK-based compression. 
Advances in SNARK proof systems have improved both general-purpose proving and specialized workloads and brought them into reality: PLONK provides succinct
verification with a universal, updatable setup~\cite{plonk},
while HyperPlonk enables linear-time proving with high-degree
custom gates~\cite{hyperplonk}.
More recently, Flock accelerates batched Boolean computations,
including standard cryptographic hashes, hash chains, and
Merkle-path verification~\cite{flock}.
Taking advantage of these developments, Succinct's Telepathy service is a prominent deployed example of SNARK proofs to compress the attestation originating with Ethereum’s 512-member ``sync committee''~\cite{telepathy}. 
A \QUASI-based alternative would improve the attestation layer feeding this pipeline; succinct proving could remain a subsequent compression step.

\paragraph{\textbf{Technical approach}}
The key idea is to distribute authentication across parties:
each party authenticates a small subset of symbols, and a quorum
of their contributions jointly attests to the pre-validated
statement. Honest parties `pin' the values of symbols they authenticate, but the challenge is to prevent corrupt parties from
changing any symbols left unauthenticated by honest parties. 


\begin{figure}[t]
\centering
\resizebox{\columnwidth}{!}{%
\begin{tikzpicture}[
  every node/.style={font=\footnotesize},
  bit/.style={
    draw=black!55,
    rounded corners=1pt,
    minimum width=6.4mm,
    minimum height=6.4mm,
    inner sep=0pt
  },
  assignment/.style={
    minimum width=6.4mm,
    minimum height=5.3mm,
    inner sep=0pt
  },
  bad/.style={
    text=red!75!black,
    font=\bfseries\footnotesize
  }
]

\node[font=\bfseries\scriptsize]
    at (1.12,0.75) {Symbol assignment};

\node[font=\bfseries\scriptsize,anchor=west]
    at (2.88,0.75) {qShare};

\node[anchor=east] at (-0.50,0) {$M$};

\foreach \j/\x/\v in {
    1/0/b_1,
    2/0.75/b_2,
    3/1.50/b_3,
    4/2.25/b_4
}{
    \node[bit,font=\scriptsize] at (\x,0) {$\v$};
}

\draw[black!25]
    (-0.85,-0.48) -- (7.8,-0.48);

\node[anchor=east]
    at (-0.50,-0.98) {$p_1$};

\foreach \x in {0,0.75,1.50}{
    \node[assignment] at (\x,-0.98)
        {$\checkmark$};
}

\node[anchor=west,font=\scriptsize]
    at (2.88,-0.98)
    {$\sigma_1=\mathsf{Sign}_{\mathsf{QOTS}}
    (b_1\parallel b_2\parallel b_3)$};

\node[anchor=east,text=red!75!black]
    at (-0.50,-1.73) {$p_2$};

\foreach \x in {0.75,1.50,2.25}{
    \node[assignment,bad]
        at (\x,-1.73) {$\times$};
}

\node[anchor=west,
      text=red!75!black,
      font=\scriptsize]
    at (2.88,-1.73)
    {$\sigma_2=\mathsf{Sign}_{\mathsf{QOTS}}
    (b_2^{*}\parallel b_3^{*}\parallel b_4^{*})$};

\node[anchor=east]
    at (-0.50,-2.48) {$p_3$};

\foreach \x in {0,1.50,2.25}{
    \node[assignment] at (\x,-2.48)
        {$\checkmark$};
}

\node[anchor=west,font=\scriptsize]
    at (2.88,-2.48)
    {$\sigma_3=\mathsf{Sign}_{\mathsf{QOTS}}
    (b_1\parallel b_3\parallel b_4)$};

\node[anchor=east]
    at (-0.50,-3.23) {$p_4$};

\foreach \x in {0,0.75,2.25}{
    \node[assignment] at (\x,-3.23)
        {$\checkmark$};
}

\node[anchor=west,font=\scriptsize]
    at (2.88,-3.23)
    {$\sigma_4=\mathsf{Sign}_{\mathsf{QOTS}}
    (b_1\parallel b_2\parallel b_4)$};

\draw[black!35]
    (-0.85,-3.65) -- (7.8,-3.65){};

\node[anchor=west,font=\scriptsize]
    at (-0.85,-4.11)
    {\textbf{Valid qAttest:}
    $(\sigma_1,\sigma_3,\sigma_4)$ --
    non-conflicting qShares};

\node[anchor=west,font=\scriptsize]
    at (-0.85,-4.67)
    {\textbf{Invalid qAttest:}
    $(\sigma_1,
    \textcolor{red!75!black}{\sigma_2},
    \sigma_3)$ --
    conflicting qShares};

\end{tikzpicture}%
}
\caption{QOTS symbol assignments and qShares
for $M=b_1 b_2 b_3 b_4$, with examples of valid and invalid
quorum attestations. Red crosses indicate altered
symbols at corrupt party $p_2$.}
\label{fig:qattest}
\end{figure}


To demonstrate this is possible, a simple deterministic construction (depicted in~\Cref{fig:qattest}) addresses this challenge
and provides modest savings:
each symbol of the statement $M$
is authenticated by $2f+1$ parties. Rotating these assignments
balances the load, so that each party authenticates approximately
two-thirds of $M$. A qShare by a party consists of a hash-based signature $\textsf{Sign}_{\mathsf{QOTS}}$ over the symbols assigned to it; any standard hash-based signatures may be employed (\Cref{sec:buildblock} elaborates on one choice).
A qAttest contains $2f+1$ valid, mutually
non-conflicting qShares, ensuring that every symbol is authenticated
by at least one honest contributor.
While this construction already reduces the per-party
contribution by one-third (asymptotically), we are confident that better
deterministic parameters are achievable and are currently
exploring improved designs.

Our main focus in this paper, however, is on larger systems,
where randomization enables more dramatic savings.
Here, \QUASI{} authenticates a codeword rather than the statement
itself. Provided the number of altered symbols remains within
the code's correction capacity, the symbols authenticated by
honest parties suffice to detect if the original pre-validated
statement has been altered. This creates a tradeoff between qShare size and
codeword expansion. \Cref{fig:quasi-random-codeword} illustrates the randomized approach
with qShares of two symbols, and a codeword with $1/3$ correcting resilience.

\paragraph{\textbf{Performance}}
Compared with a naïve collection of hash-based signatures,
\QUASI{} targets savings of more than $10\times$ in
(i) per-party contribution size, (ii) generation effort,
and (iii) verification effort.
These gains arise from reducing the number of hash-chain values
and preimages maintained or revealed by each party, as well as
the hash computations required for generation and verification.
For a 256-bit statement digest $M$, \QUASI{} represents the input
as $m=32$ symbols over 
$\mathrm{GF}(2^8)$ and employs an
error-correcting code with $3\times$ expansion.
It sets the number of symbols authenticated per qShare, denoted $d$,
to ensure that undetected symbol flipping occurs with
probability less than $2^{-128}$.
Table~\ref{tab:d-vals} shows a selection of concrete parameters
and the resulting savings.

The overall goal is to present the defense against the strongest adversary with the strongest security parameters. However, alongside we present a common case, which may have proportionally fewer amount of Byzantine failures. Hence, we consider possible pragmatic variations which include proportionally reducing number of Byzantine failures by defining $f = b+c$ ($b$ are Byzantine and $c$ crash failures).

\paragraph{\textbf{Operations}}
QUASI requires no additional interactive setup or preprocessing
beyond key generation for the underlying hash-based authentication
scheme. Parties generate keys independently for $d$ symbols
rather than $m$, and publish their selected symbol positions
alongside their verification keys. Each party then generates
its qShare using $\textsf{Sign}_{\mathsf{QOTS}}$.
Anyone can collect $2f+1$ valid, mutually non-conflicting qShares
on $M$ to form a qAttest.

Across multiple QUASI instances, parties can manage keys using
established techniques from stateful hash-based signature schemes,
such as XMSS~\cite{xmss}, which manages one-time WOTS-family keys.
Key management is orthogonal to the QUASI construction and
outside the scope of this work.

\section{Preliminaries}
\label{sec:prelims}

This work introduces a newly conceived class of attestation protocols named \QUASI. 
We first give an informal description of \QUASI, and follow up more formally in~\Cref{def:primitive}. 
We also present a glossary of notations and terms used throughout the construction in~\Cref{tab:glossary}.

\begin{table}[t]
\caption{Glossary of notation.}
\label{tab:glossary}
\centering
\scriptsize
\setlength{\tabcolsep}{3pt}
\renewcommand{\arraystretch}{1.4}
\arrayrulecolor{black!35}

\renewcommand{\tabularxcolumn}[1]{m{#1}}

\begin{tabularx}{\columnwidth}{
@{}
>{\centering\arraybackslash}m{0.22\columnwidth}
>{\centering\arraybackslash}m{0.34\columnwidth}
>{\centering\arraybackslash}X
@{}
}
\hline
\rowcolor{gray!20}
\textbf{Term} & \textbf{Notation} & \textbf{Meaning} \\
\hline

Statement
& $M$, $m$
& Input to be certified, containing $m$ symbols \\
\hline

\rowcolor{gray!8}
Encoded statement
& $c(M)$, $\hat{m}$
& ECC with expansion
  $\lambda=\hat{m}/m$ \\
\hline

Symbol space
& $\mathcal{A}$
& Alphabet of 8-bit values \\
\hline

\rowcolor{gray!8}
Symbol
& $b_j$, $c_j$, $M|_j$, $c(M)|_j$
& The $j$-th symbol of $M$ or $c(M)$ \\
\hline

QUASI-OTS
& $\textsf{Sign}_{\mathsf{QOTS}}(w)$
& One-time hash-based signature over a word $w$ \\
\hline

\rowcolor{gray!8}
qShare map
& $\pi_i:[d]\rightarrow[\hat{m}]$, $d$
& Party $i$'s symbol assignment \\
\hline

qShare
& $\sigma_i(M)$
& Party $i$'s quasi signature-share over $M$ \\
\hline

\rowcolor{gray!8}
qAttest
& $\sigma(M)$
& Full attestation over $M$ \\
\hline

\end{tabularx}
\arrayrulecolor{black}
\end{table}

In \QUASI, a group of $n=3f+1$ parties, up to $f$ of which may be corrupt,
can produce an attestation about a pre-validated statement
on behalf of quorums of $\geq 2f+1$ parties, 
and no such quorum can forge an attestation, which the honest parties in the quorum did not intend to provide, except with negligible probability\footnote{We set the security guarantee at a conventional level $1-2^{-128}$.}. 

Parties in the \QUASI primitive each hold an independently generated private key for signing, similar to multi-signature schemes~\cite{MicaliOhtaReyzin01}; the public keys of all parties are known to everyone. 
Each party contributes a quasi-signature share (qShare) on a statement, which does not constitute by itself a valid signature on the statement,
akin to classical threshold signature schemes~\cite{DesmedtFrankel91}.
A quorum of qShares produces a joint quorum attestation (qAttest), which is not required to look like a single signature. Furthermore, there are no secrecy guarantees, and a qAttest may reveal and hold accountable who contributed to an attestation, similar to aggregate signature paradigms~\cite{BonehGentryLynnShacham03}. 


However, while being inspired by signature aggregation schemes, \QUASI is not one. It is designed for quorum-based systems to attest to a pre-validated statement,
hence it provides a weaker unforgeability guarantee, called \textit{weak{-}$f${-}unforgeability}, defined below.
The goal is to attest to external parties that the statement is valid.

\medskip

\noindent More formally, a \QUASI scheme is defined as follows.

\begin{definition}[\textbf{\QUASI}]
\label{def:primitive}
A \emph{\QUASI} scheme is a tuple of probabilistic polynomial-time
algorithms
$\Pi = (\textsf{qSetup},\, \textsf{qKeyGen},\, \textsf{qShareSign},\,
         \textsf{qCombine},\, \textsf{qVerify})$
involving $n$ parties $p_1, \ldots, p_n$ and a threshold
$f$ with $1 \le f \le n$, defined as follows.

\begin{itemize}

\item $\textsf{qSetup}(1^\kappa, n, f)\to \mathsf{pp}$.
  On input a security parameter $\kappa$, and environment parameters $n$ and $f$, outputs public parameters
  $\mathsf{pp}$ shared by all parties, including:
  a description of the statement space $\mathcal{M}$; 
  an error-correcting code  $c(\cdot)$ over $\mathcal{M}$ which includes an expansion parameter $\lambda$;  
  the quasi signature share (qShare) space $\Sigma_{\mathsf{sh}}$ which includes a qShare size parameter $d$;
  and the quorum attestation (qAttest) space $\Sigma$.

\item $\textsf{qKeyGen}(\mathsf{pp}) \to (\mathsf{vk}_i,\, \mathsf{sk}_i)$.
  A randomized algorithm run \emph{independently} by each party
  $p_i$; outputs a verification key $\mathsf{vk}_i$ and a signing key
  $\mathsf{sk}_i$.
  The combined verification key is
  $\mathsf{VK} = (\mathsf{vk}_1, \ldots, \mathsf{vk}_n)$.

\item $\textsf{qShareSign}(\mathsf{pp},\, \mathsf{sk}_i,\, M) \to \sigma_i$.
  On input the public parameters, party $p_i$'s signing key, and a
  statement $M \in \mathcal{M}$, outputs a \emph{quasi signature share} (qShare)
  $\sigma_i \in \Sigma_{\mathsf{sh}}$.
  A qShare does not constitute a valid signature over $M$ on its own.

\item $q\textsf{Combine}(\mathsf{pp},\, \mathsf{VK},\,
                        S,\, M,\, \{\sigma_i\}_{i \in S}) \to \sigma
                        \;\text{or}\; \bot$.
On input the public parameters, the combined verification key, an
  index set $S \subseteq [n]$ with $|S| \ge 2f+1$, a statement $M$, and
  the qShares $\{\sigma_i\}_{i \in S}$, outputs either a
  qAttest $\sigma \in \Sigma$ or a failure symbol $\bot$.

\item $\textsf{qVerify}(\mathsf{pp},\, \mathsf{VK},\, M,\, \sigma)
      \to \{0, 1\}$.
  On input the public parameters, the combined verification key, a
  statement $M$, and a purported qAttest $\sigma$, outputs
  $1$ (accept) or $0$ (reject).

\end{itemize}

\noindent
We require the following properties.

\paragraph{\textbf{Correctness}}
For all $\mathsf{pp} \leftarrow \textsf{qSetup}(1^\kappa, n, f)$, all
$(\mathsf{vk}_i, \mathsf{sk}_i) \leftarrow \textsf{qKeyGen}(\mathsf{pp})$,
all statements $M \in \mathcal{M}$, all index sets $S \subseteq [n]$
with $|S| \ge 2f+1$, and all shares
$\sigma_i \leftarrow \textsf{qShareSign}(\mathsf{pp}, \mathsf{sk}_i, M)$
for $i \in S$:\[
  \Pr\left[
    \begin{array}{l}
      \sigma \leftarrow \textsf{qCombine}(\mathsf{pp},\mathsf{VK},S,M,
        \{\sigma_i\}_{i\in S}) :\\[2pt]
      \textsf{qVerify}(\mathsf{pp},\mathsf{VK},M,\sigma) = 1
    \end{array}
  \right] = 1
\] 


\paragraph{\textbf{\unforge}}
For all $\mathsf{pp} \leftarrow \textsf{qSetup}(1^\kappa, n, f)$, all
$(\mathsf{vk}_i, \mathsf{sk}_i) \leftarrow \textsf{qKeyGen}(\mathsf{pp})$,
all probabilistic polynomial-time adversaries $\mathcal{A}$
controlling a coalition $\mathcal{C} \subset [n]$ of corrupted parties
with $|\mathcal{C}| \leq f$, having access to a signing oracle $O^M$ which is allowed to make $qShareSign(\mathsf{pp}, \mathsf{sk}_i, M)$ requests only on a pre-validated $M$: \begin{multline*}
  \Pr\Big[\,
    \textsf{qVerify}(\mathsf{pp},\mathsf{VK},M^*,\sigma^*) = 1
    \;\Big|\; M^* \neq M, \\
    \sigma^* \leftarrow \mathcal{A}^{\textsf{O}^M}\big(\mathsf{pp},\mathsf{VK},
      \{\mathsf{sk}_i\}_{i\in\mathcal{C}}\big)
  \,\Big] \le \mathsf{negl}(\kappa).
\end{multline*}


\paragraph{\textbf{Transparency}}
The quorum attestation $\sigma$ output by \textsf{qCombine} may
encode the signing set $S$ and is not required to be
indistinguishable from a single-signer signature.
In particular, no anonymity guarantee is imposed on
$\sigma$ with respect to the identities of the signers in $S$.

\paragraph{\textbf{Remarks}}
It is worthy of noting how \unforge differs from traditional threshold-unforgeability: in the latter, the adversary 
has access to a signing oracle $O^{\mathcal{Q}}$ which is allowed to make $qShareSign(\mathsf{pp}, \mathsf{sk}_i, M')$ requests on any $M' \in \mathcal{Q}$. That is, it allows the adversary to query honest parties on a set  $\mathcal{Q}$ of statements, rather than a unique statement $M$. It then requires negligible forging success probability for all statements $M^* \notin \mathcal{Q}$, instead of $M^* \neq M$, not queried to any honest party. 

\QUASI is not designed to withstand this form of attack. To shed insight on this, recall that each party authenticates only $d$ codeword positions. There exist two distinct statements $M, M^*$ whose encodings agree at every position selected by honest party $p_h$. In other words, both codewords have identical symbols at all positions in $\pi_h$, which are authenticated by $p_h$.
If the oracle is allowed to query honest parties on both $M$ and $M^*$, then  
the adversary can form a valid qAttest on $M$ by combining qShares from $p_h$, a set $A$ of $f$ honest parties, and a set $B$ of $f$ corrupt parties; and form another valid qAttest on $M^*$ with qShares from $p_h$, a distinct set $A'$ of $f$ honest parties, and $B$.

\paragraph{\textbf{Operations}}
QUASI requires no special protocol or preprocessing.
For each context, parties independently generate the underlying
($\textsf{Sign}_{\mathsf{QOTS}}$) keys material, but for far fewer symbols, without
coordination or trusted setup. Each public key additionally
specifies the $d$ selected symbol positions.
To contribute a qShare on $M$, a party applies the underlying
authentication scheme to the symbols at those positions.
Any entity may collect contributions from a set
$S\subseteq[n]$, $|S|=2f+1$, and form
$\sigma(M)=\{(i,\sigma_i(M)):i\in S\}$,
provided that the qShares are valid and mutually
non-conflicting.

\paragraph{\textbf{Stateful key management}}
Across multiple QUASI instances, parties can manage keys using established techniques from stateful hash-based signature schemes, such as XMSS~\cite{xmss}, which supports the management of one-time WOTS-family keys. We treat key management as orthogonal to the QUASI construction and leave it outside the scope of this work. 

\end{definition}


\subsection{Adversary models}

We consider two possible adversary models: \emph{fully adaptive} and \emph{semi-adaptive}. Both models allow the adversary to corrupt up to $f$ parties, but differ in when it can select which parties to corrupt. In the \emph{fully adaptive} adversary model, the adversary can select $f$ parties to corrupt at any point during the protocol execution.
The \emph{semi-adaptive} adversary model requires the adversary to commit to a specific set of at most $f$ parties before the protocol begins. In both models, the adversary can then interact with the remaining $2f+1$ honest parties through the oracle that provides information about their qShares. The following sections prioritize security against a \emph{fully adaptive} adversary, however also considers the \emph{semi-adaptive} model, which enables stronger efficiency advancements. We will also consider more practical cases, assuming a smaller proportion of Byzantine failures $f = b +c$, where $b$ is the maximal number of Byzantine faults, and $c$ additional crash faults. 


\subsection{Performance Measure}

Our primary performance measure is the number of input bits
authenticated per party, or equivalently, the number of symbols
when the symbol size is fixed. For a statement $M$ containing
$m$ symbols, we measure the reduction from authenticating all
$m$ symbols to authenticating only $d$ symbols per qShare,
giving a reduction factor of $m/d$.

The following costs are all derived directly from this measure: the authentication storage required for
a qAttest containing $2f+1$ qShares, the communication cost of
collecting those qShares and responding to client queries, and
the hash computations required to generate and verify each
party's contribution. The per $d$-symbol authentication costs are inherited directly from the underlying authentication method,
such as WOTS (\Cref{sec:buildblock}); the savings arise from
reducing the number of input symbols it authenticates.

These reductions also benefit SNARK-based aggregation.
The attestation size and the hash computations required for
verification contribute directly to the input size and
verification work represented within the proof. Reducing
these quantities relative to a collection of full hash-based
signatures therefore yields proportional savings in the
corresponding components of the proving workload.

\section{Building Blocks} \label{sec:buildblock}

\subsection{QUASI-OTS}

We employ an underlying hash-based signature scheme for $d$-symbol authentication.
Each key pair is used for at most one signing invocation.

Let $\mathcal{A}=\{0,1\}^{8}$ be the symbol alphabet.
QUASI-OTS is a hash-based one-time signature construction for
words of $d$ symbols, using WOTS to authenticate the concatenated
$8d$-bit word, including its WOTS checksum.
Its standard signing interface is

\begin{equation*}
\begin{aligned}
\Pi_{\mathsf{QOTS}} = (&
\textsf{Setup}_{\mathsf{QOTS}},
\textsf{KeyGen}_{\mathsf{QOTS}},\\
&
\textsf{Sign}_{\mathsf{QOTS}},
\textsf{Verify}_{\mathsf{QOTS}}).
\end{aligned}
\label{eq:qots}
\end{equation*}

\begin{itemize}
\item
$\textsf{Setup}_{\mathsf{QOTS}}(1^\kappa,d)
 \to\mathsf{pp}_{\mathsf{QOTS}}$.
Output public parameters specifying the security parameter
$\kappa$, the word space $\mathcal{A}^{d}$, the WOTS
instantiation and its parameters, and the signing-key,
verification-key, and signature spaces.

\item
$\textsf{KeyGen}_{\mathsf{QOTS}}(\mathsf{pp}_{\mathsf{QOTS}})
 \to(\mathsf{vk}^{\mathsf{QOTS}},\mathsf{sk}^{\mathsf{QOTS}})$.
Generate a WOTS key pair for authenticating words in
$\mathcal{A}^{d}$.

\item
$\textsf{Sign}_{\mathsf{QOTS}}
 (\mathsf{pp}_{\mathsf{QOTS}},\mathsf{sk}^{\mathsf{QOTS}},a)
 \to\tau$.
For $a=a_1\cdots a_d\in\mathcal{A}^{d}$, concatenate the
symbols in their given order and apply WOTS encoding and
signing, including the checksum.
Output the resulting signature $\tau$.

\item
$\textsf{Verify}_{\mathsf{QOTS}}
 (\mathsf{pp}_{\mathsf{QOTS}},\mathsf{vk}^{\mathsf{QOTS}},a,\tau)
 \to\{0,1\}$.
Reject malformed inputs. Otherwise, apply WOTS verification
to the same ordered word and its checksum.
\end{itemize}

QUASI-OTS inherits correctness and one-time unforgeability
from the underlying WOTS instantiation.
For completeness, we elaborate on the details of the underlying WOTS signing scheme in Appendix~\ref{app:qots}.

\subsection{Expanding Input to a Codeword}
We employ a Reed-Solomon (RS) error-correcting code (ECC), denoted $c(\cdot)$, over statements $M \in \mathcal{M}$ (other codes may be substituted if needed).

A statement $M$ is represented as a vector of $m=|M|/w$ symbols over an alphabet of $w$-bit symbols. 
We need to make sure that symbol width $w$ is sufficiently large.
The reason is that RS employs polynomials over $GF(2^w)$. In order to be able to encode $m$ interpolation points, the field GF($2^w$) must have size at least $m$, i.e., $2^w \geq m$. In our case, usually the statement to sign is a cryptographic digest of length $256$ bits.
The smallest $w$ that satisfies $2^w \geq 256$ is $w=8$. Therefore, henceforth $M$ is represented by $m = 32$ symbols, each of $8$ bits. 

With a code of size $\hat{m} = |c(M)| = \lambda m$, an RS code can correct $e$ corruptions in unknown symbols as well as $o$ omissions of known symbols, such that $2e + o \le (\lambda-1)m$. Put differently, every pair of codewords differ in at least $\dmin \coloneqq (\lambda-1)m+1$ symbols.
The factor $\lambda = \hat{m}/m$ is called the expansion factor. 


\section{The \QUASI Scheme} \label{sec:encoded}

\begin{figure*}[t]
\centering
\resizebox{\textwidth}{!}{%
\begin{tikzpicture}[
  every node/.style={font=\scriptsize},
  bit/.style={draw=black!65,minimum width=6mm,minimum height=6mm,inner sep=0pt},
  cslot/.style={draw=black!55,fill=black!7,minimum width=6.7mm,
                minimum height=6mm,inner sep=0pt},
  selected/.style={draw=blue!65!black,rounded corners=1pt,
                   minimum width=6.2mm,minimum height=5.8mm,inner sep=0pt},
  changed/.style={draw=red!75!black,text=red!75!black,
                  rounded corners=1pt,minimum width=6.2mm,
                  minimum height=5.8mm,inner sep=0pt},
  redtext/.style={text=red!75!black}
]
\def\sx{2.30}
\def\dx{0.78}

\node[anchor=east] at (5.12,3.40) {$M$};

\foreach \i in {1,2,3,4} {
  \node[bit] at ({5.52+(\i-1)*0.72},3.40) {$b_{\i}$};
}

\draw[-{Stealth[length=2mm]},thick]
  (6.60,3.02) -- node[right,font=\scriptsize] {encode} (6.60,2.15);

\node[anchor=east] at (1.88,1.62) {Codeword $c(M)$};
\foreach \i in {1,...,12} {
  \node[cslot] at ({\sx+(\i-1)*\dx},1.62) {$c_{\i}$};
}
\draw[black!30] (1.50,1.08) -- (18.90,1.08);

\node[anchor=east,font=\bfseries\scriptsize] at (1.88,0.69) {Party};
\node[font=\bfseries\scriptsize] at (6.57,0.69)
  {Symbol assignment $\pi_i$};
\node[anchor=west,font=\bfseries\scriptsize] at (11.58,0.69) {qShare};

\newcommand{\honestmark}[2]{%
  \node[selected] at ({\sx+((#1)-1)*\dx},#2) {$\checkmark$};%
}
\newcommand{\redmark}[2]{%
  \node[changed] at ({\sx+((#1)-1)*\dx},#2) {$\checkmark$};%
}
\newcommand{\crossmark}[2]{%
  \node[changed] at ({\sx+((#1)-1)*\dx},#2) {$\times$};%
}

\node[anchor=east] at (1.88,0.05) {$p_1$};
\honestmark{2}{0.05}\honestmark{6}{0.05}
\node[anchor=west] at (11.58,0.05)
  {$\sigma_1=\mathsf{Sign}_{\mathsf{QOTS}}(c_2\parallel c_6)$};

\node[anchor=east] at (1.88,-0.71) {$p_2$};
\honestmark{5}{-0.71}\honestmark{9}{-0.71}
\node[anchor=west] at (11.58,-0.71)
  {$\sigma_2=\mathsf{Sign}_{\mathsf{QOTS}}(c_5\parallel c_9)$};

\draw[blue!65!black,densely dashed,line width=.55pt]
  ({\sx+4*\dx},-1.01) -- ({\sx+4*\dx},-1.17);

\node[anchor=east,redtext] at (1.88,-1.47) {$p_3$ (corrupt)};
\redmark{5}{-1.47}\crossmark{8}{-1.47}
\node[anchor=west,redtext] at (11.58,-1.47)
  {$\sigma_3=\mathsf{Sign}_{\mathsf{QOTS}}(c_5\parallel c_8^*)$};

\node[anchor=east] at (1.88,-2.23) {$p_4$};
\honestmark{1}{-2.23}\honestmark{11}{-2.23}
\node[anchor=west] at (11.58,-2.23)
  {$\sigma_4=\mathsf{Sign}_{\mathsf{QOTS}}(c_1\parallel c_{11})$};

\node at (1.74,-2.98) {$\vdots$};
\node at (6.58,-2.98) {$\vdots$};
\node at (14.0,-2.98) {$\vdots$};

\node[anchor=east] at (1.88,-3.73) {$p_{n-1}$};
\honestmark{3}{-3.73}\honestmark{12}{-3.73}
\node[anchor=west] at (11.58,-3.73)
  {$\sigma_{n-1}=\mathsf{Sign}_{\mathsf{QOTS}}(c_3\parallel c_{12})$};

\node[anchor=east] at (1.88,-4.49) {$p_n$};
\honestmark{4}{-4.49}\honestmark{7}{-4.49}
\node[anchor=west] at (11.58,-4.49)
  {$\sigma_n=\mathsf{Sign}_{\mathsf{QOTS}}(c_4\parallel c_7)$};

\draw[black!30] (1.50,-4.95) -- (18.90,-4.95);


\node[anchor=west,text opacity=0] at (-2.00,-5.42)
  {\textbf{Invalid qAttest:} $(\sigma_1,\sigma_2,\textcolor{red!75!black}{\sigma_3},\ldots,\sigma_{2f+1})$};

\node[anchor=west,text opacity=0] at (3.50,-5.42)
  {\textbf{Codeword check:} $\tilde c=(c_1,\ldots,c_7,c_8^*,c_9,\ldots,c_{12})$, where $c_8$ is detected};

\node[anchor=center,overlay,font=\small] at (8.45,-5.42)
  {\textbf{Invalid qAttest:}
   $(\sigma_1,\sigma_2,\textcolor{red!75!black}{\sigma_3},
   \ldots,\sigma_{2f+1})$
   \qquad
   \textbf{Codeword check:}
   $\tilde c=(c_1,\ldots,c_7,c_8^*,c_9,\ldots,c_{12})$ fails};

\end{tikzpicture}%
}
\caption{Randomized symbol assignments and QOTS shares for the codeword
$c(M)$ of $M=b_1b_2b_3b_4$ . Each party authenticates two symbols.
The honest party $p_2$ and corrupt party $p_3$ both authenticate $c_5$,
whereas only $p_3$ authenticates $c_8$. The alteration $c_8^*$ creates
no pairwise authentication conflict but is detected by the codeword check.}
\label{fig:quasi-random-codeword}
\end{figure*}
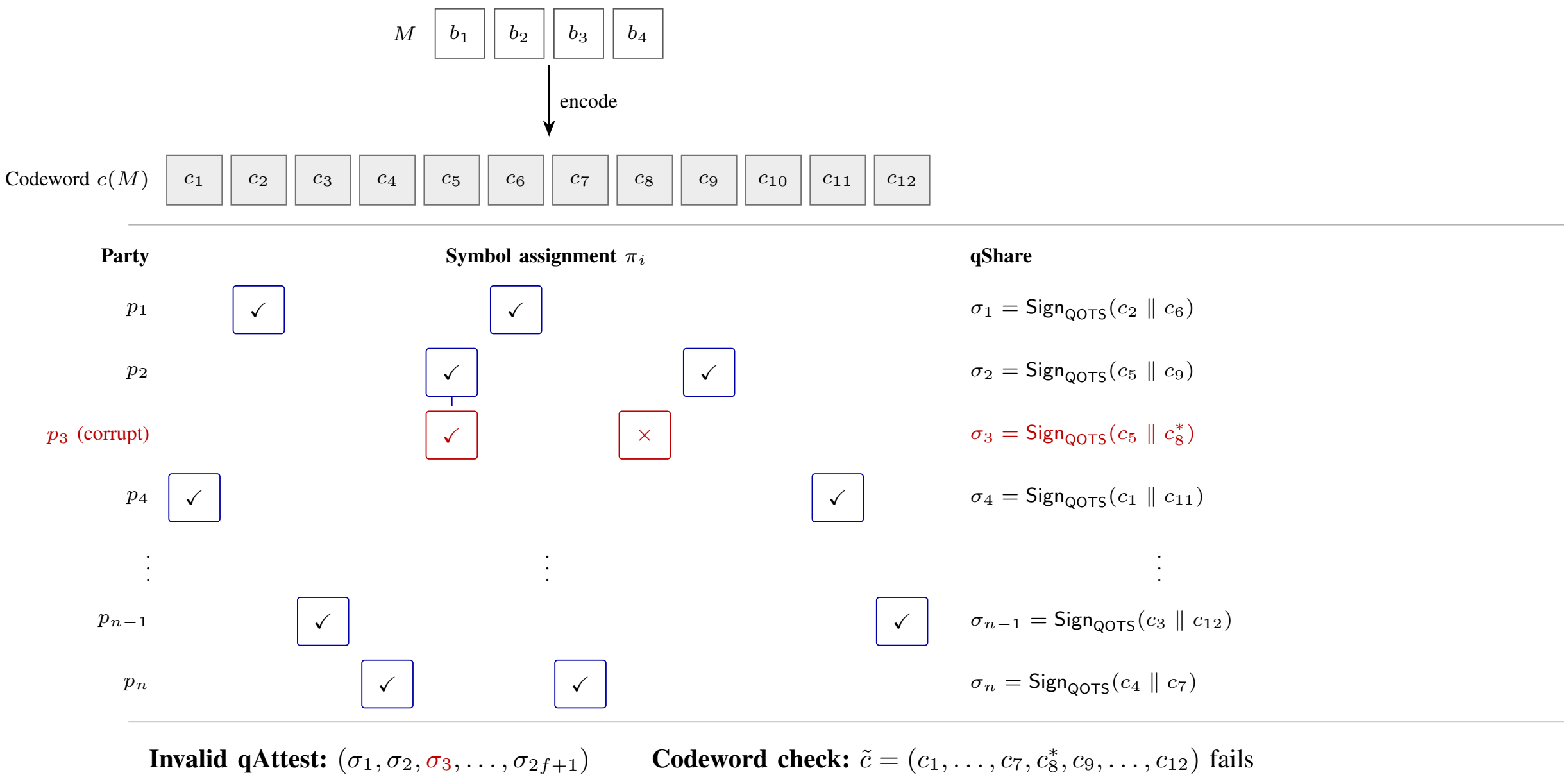

In this section, we develop the foundations of a \QUASI construction with an ambitious goal: each party authenticates only $d$ symbols, for a small constant $d$ chosen as a function of $n$.

It is easy to see that unless each symbol of $M$ is authenticated by $2f{+}1$ parties, there exists a collection of $2f{+}1$ parties and a symbol $b^* = M{\mid}_\ast$ such that $b^*$ is not authenticated by any honest party.
We refer to such a $b^*$ as \textit{flipped}: 
an adversary could produce a qAttest mixing authenticators over valid $M$ symbols by honest parties with authenticators only by corrupt parties over $b^*$ flipped.

In order to nevertheless get below $2f{+}1$ authenticators per $M$ symbol, \QUASI authenticates a codeword $c(M)$ over $M$, rather than $M$ itself. 
For example, ~\Cref{fig:quasi-random-codeword} illustrates the \QUASI approach with qShares of $d=2$ symbols over an encoding $c(M)$ with expansion $\lambda = 3$.

Next, we describe the recipe for parties to authenticate subsets of $c(M)$.

\subsection{Generic QUASI Construction}
The family of QUASI constructions separates three interchangeable components:
the encoding $c$, the symbol-assignment regime, and the
number $d$ of symbols authenticated per party.
Generally, the assignment may be deterministic or randomized; this paper focuses on the latter.
Their common requirement is a coverage condition:
every set of $f+1$ honest contributors must leave at most
$2e$ \emph{`holes'}, for the code that has minimum distance $\dmin = 2e+1$.

\paragraph{\textbf{Parameters and Coverage}}

Let $n = 3f+1$, with at most $f$ corrupt parties, and let
a qAttest contain $2f+1$ qShares.
Choose an injective encoding
$c:\mathcal{A}^{m}\to\mathcal{A}^{\hat m}$
as described in ~\Cref{sec:buildblock}, with minimum Hamming distance at least $2e+1$.
Its expansion factor is $\lambda=\hat m/m$.

An assignment regime specifies a public mapping $\pi_i:[d]\to[\hat m]$ for each party $p_i$.
The mappings are fixed during set up time. 
For $H\subseteq[n]$, define its covered positions and holes by:\[
U(H)\coloneqq
\bigcup_{i\in H}\pi_i([d]),
\qquad
E(H)\coloneqq[\hat m]\setminus U(H).
\]
The required coverage condition is: \begin{equation*}    
\forall H\subseteq[n],\quad
|H|=f+1
\ \Longrightarrow\
|E(H)|\leq2e.
\label{eq:quasi-coverage}
\end{equation*}

For deterministic assignments, this condition holds
unconditionally.
For randomized assignments, the analysis bounds the
probability that it fails.
For an MDS code, including Reed--Solomon,
$\hat m=m+2e$, so the condition is equivalent to
$|U(H)|\geq m$.

\paragraph{\textbf{API}}

The construction uses QUASI-OTS to authenticate an ordered
word of $d$ symbols.
Define the word selected by party $p_i$ as:
\[
a_i(M)\coloneqq
c(M)\mid_{\pi_i(1)}
\parallel\cdots\parallel
c(M)\mid_{\pi_i(d)}.
\]

\paragraph{$\textsf{qSetup}(1^\kappa,n,f)\to\mathsf{pp}$}
Check that $n\geq3f+1$; otherwise, output $\bot$.
Select the statement space $\mathcal{M}=\mathcal{A}^{m}$,
the encoding $c$, the parameter $d$, and an assignment
regime specifying $(\pi_1,\ldots,\pi_n)$.
Run
$\mathsf{pp}_{\mathsf{QOTS}}
\leftarrow\textsf{Setup}_{\mathsf{QOTS}}(1^\kappa,d)$.
The choices of $c$, $d$, and the assignment regime must
satisfy~\eqref{eq:quasi-coverage}, or provide a bound on
its failure probability.

Output public parameters specifying
$n$, $f$, $\mathcal{M}$, $c$, $\hat m$, $e$, $d$,
the public mappings, and
$\mathsf{pp}_{\mathsf{QOTS}}$.
Let $\Sigma_{\mathsf{sh}}$ be the QUASI-OTS signature space,
and define the qAttest space as: \[
\Sigma\coloneqq
\left\{
  \{(i,\sigma_i):i\in S\}
  \;\middle|\;
  S\subseteq[n],\ |S|=2f+1,\
  \sigma_i\in\Sigma_{\mathsf{sh}}
\right\}.
\]

\paragraph{$\textsf{qKeyGen}(\mathsf{pp})
\to(\mathsf{vk}_i,\mathsf{sk}_i)$}
Party $p_i$ independently generates:\[
(\mathsf{vk}^{\mathsf{QOTS}}_i,
 \mathsf{sk}^{\mathsf{QOTS}}_i)
\leftarrow
\textsf{KeyGen}_{\mathsf{QOTS}}
(\mathsf{pp}_{\mathsf{QOTS}}).
\]
Set
$\mathsf{vk}_i=(\pi_i,\mathsf{vk}^{\mathsf{QOTS}}_i)$
and
$\mathsf{sk}_i=(\pi_i,\mathsf{sk}^{\mathsf{QOTS}}_i)$.
The registered verification keys form
$\mathsf{VK}=(\mathsf{vk}_1,\ldots,\mathsf{vk}_n)$.

\paragraph{$\textsf{qShareSign}
(\mathsf{pp},\mathsf{sk}_i,M)\to\sigma_i$}
For a pre-validated statement $M\in\mathcal{M}$,
party $p_i$ authenticates its selected word:
\[
\sigma_i\leftarrow
\textsf{Sign}_{\mathsf{QOTS}}
\bigl(
  \mathsf{pp}_{\mathsf{QOTS}},
  \mathsf{sk}^{\mathsf{QOTS}}_i,
  a_i(M)
\bigr).
\]
Each QUASI-OTS key pair is used for at most one
$\textsf{qShareSign}$ invocation.

\paragraph{$\textsf{qCombine}
(\mathsf{pp},\mathsf{VK},S,M,\{\sigma_i\}_{i\in S})
\to\sigma\text{ or }\bot$}
Require $S\subseteq[n]$ and $|S|=2f+1$.
Form $\sigma=\{(i,\sigma_i):i\in S\}$ and output it
if $\textsf{qVerify}(\mathsf{pp},\mathsf{VK},M,\sigma)=1$;
otherwise, output $\bot$.

\paragraph{{$\textsf{qVerify}
(\mathsf{pp},\mathsf{VK},M,\sigma)\to\{0,1\}$}}
Reject malformed inputs, statements outside $\mathcal{M}$,
and attestations not containing exactly $2f+1$ distinct
registered parties.
For every $(i,\sigma_i)\in\sigma$, compute $a_i(M)$ and check:
\[
\textsf{Verify}_{\mathsf{QOTS}}
\bigl(
  \mathsf{pp}_{\mathsf{QOTS}},
  \mathsf{vk}^{\mathsf{QOTS}}_i,
  a_i(M),
  \sigma_i
\bigr)=1.
\]
Accept if and only if all checks pass.
Thus, every accepted qShare authenticates symbols of the
same claimed codeword $c(M)$, and the qShares are mutually
non-conflicting.
Verification requires exact agreement with the claimed
codeword; it does not correct conflicting authenticated
values.

\subsection{Correctness and Security}

\begin{theorem}[Validity]
\label{thm:validity}
For every $M\in\mathcal{M}$ and every set $S\subseteq[n]$
of $2f+1$ honest parties, let
$\sigma_i\leftarrow
\textsf{qShareSign}(\mathsf{pp},\mathsf{sk}_i,M)$
for each $i\in S$.
Then $\textsf{qCombine}$ outputs a qAttest $\sigma$ satisfying:\[
\textsf{qVerify}(\mathsf{pp},\mathsf{VK},M,\sigma)=1
\]
with probability $1$, independently of the coverage condition.
\end{theorem}

\begin{proof}
By correctness of QUASI-OTS, every $\sigma_i$ passes
verification on the selected word $a_i(M)$.
All contributions authenticate symbols of the same codeword
$c(M)$ and are therefore mutually non-conflicting.
Since $S$ contains $2f+1$ distinct registered parties,
$\textsf{qCombine}$ and $\textsf{qVerify}$ accept.
\end{proof}

\begin{theorem}[Weak-$f$-unforgeability]
\label{thm:unforge}
Suppose the coverage condition in~\Cref{eq:quasi-coverage}
holds and the code has minimum distance
$\dmin=2e+1$.
In a context where honest parties contribute only on the
pre-validated statement $M$, an adversary controlling at
most $f$ parties cannot produce an accepted qAttest for
$M'\ne M$, except with the probability of breaking the
underlying QUASI-OTS authentication.
\end{theorem}

\begin{proof}
An accepted qAttest contains contributions from $2f+1$
distinct parties, including at least $f+1$ honest parties.
Choose a set $H$ of $f+1$ such honest contributors.

Unless QUASI-OTS authentication is broken, each accepted
honest contribution authenticates the same selected word
as its contribution on $M$.
An accepted contribution from an honest party that never
contributed in this context would itself constitute an
authentication forgery.
Consequently,
$c(M')\mid_j=c(M)\mid_j$ for every $j\in U(H)$.

The two codewords can therefore differ only at holes:
\[
\operatorname{dist}_{\mathrm{H}}\bigl(c(M),c(M')\bigr)
\leq |E(H)|
\leq2e.
\]
Distinct codewords have distance at least $\dmin=2e+1$.
Hence $c(M')=c(M)$, and injectivity of $c$ gives $M'=M$.
\end{proof}

If a randomized assignment violates the coverage condition
with probability at most $\epsilon_{\mathsf{assign}}$,
the overall forgery probability is at most
$\epsilon_{\mathsf{assign}}+\epsilon_{\mathsf{auth}}$,
where $\epsilon_{\mathsf{auth}}$ bounds the probability
of a successful attack on the underlying authentication.
Deterministic constructions eliminate the assignment-failure
term.

\section{The Randomized Construction} 
\label{sec:rand}

We now present the randomized QUASI variant instantiating the scheme in~\Cref{sec:encoded}. The main benefit of randomizing symbol selection is a graceful degradation against varying degrees of actual corrupt parties.

Fix $m=32$ eight-bit statement symbols and a Reed--Solomon
encoding of length $\hat m=96$ and minimum distance
$\dmin = 65 = 2e+1$, where $e=32$.
During setup, each party independently samples a mapping $\pi_i:[d]\to[96]$,
choosing each $\pi_i(j)$ independently and uniformly at random, without repetition.

Choosing the number $d$ of symbols each party authenticates from $c(M)$ 
creates a tradeoff between the size of per-party qShare contribution and amount of potentially flipped symbols. 
Provided the number of flipped symbols is bounded accordingly, the valid symbols which are authenticated by honest parties suffice to detect and prevent these flips. 
The larger the system, the smaller is a sufficient $d$.
\Cref{sec:bounds} provides an analysis proving that flipped symbols remain within the code's correction capacity with probability $1-2^{-128}$ for suitable choices of $d$, both against a fully-adaptive adversary and a semi-adaptive one; a sample of concrete values is then calculated in~\Cref{sec:concrete}.

\subsection{Forgery Probability Bounds} 
\label{sec:bounds}

We proceed with the analysis of the relationship between the QUASI parameters that are needed to achieve negligible forgery probability, under different adversarial models and security settings. 
The analysis below links between $n=3f{+}1$, the environment; and $d$, the choice of qShare size.

The adversary game we analyze is modeled as follows: 
honest parties choose their mappings $[d] \rightarrow [\hat{m}]$ uniformly at random and publish them along with their public QUASI-OTS keys. 

The adversary can select any subset $S'$ of $f{+}1$ honest parties and query for their qShares over $M$. It can use $f$ corrupt parties under its control to generate their qShares over $c(M)^*$ and form a qAttest, where $c(M)^*$ has certain symbols flipped at positions left unauthenticated by $S'$.  
The adversary wins if the fraction of the symbols in $c(M)$ that $S'$ leaves unauthenticated matches the code distance $\dmin$, allowing the adversary to undetectably flip symbols.

For each subset $S'$ of $f{+}1$, 
denote $E(S')$ the total number of symbols left unauthenticated by $S'$.
The game's outcome is the maximum size of $E(S')$ over all selections of sets $S'$:
$e_{actual} := \max \{E(S') \mid |S'|=f{+}1 \}$. The goal is to link $d$, $n$, and $f$, such that with high probability $e_{actual} < \dmin$.

\paragraph{\textbf{Adaptive adversary}}
We proceed with a probabilistic analysis with a fully adaptive adversary, who may select $f$ parties to corrupt after seeing the honest parties' mappings. 
Abstractly, this game consists of a set of $3f{+}1$ \textit{throws} $\{\pi_i\}_{i=1..(3f{+}1)}$, representing the mappings $\pi_i$ selected by parties before the adversary chooses $f$ to corrupt. Every $\pi_i$ consists of $d$ balls. Each throw places $d$ balls into $\hat{m}$ bins uniformly at random, such that no two balls in a throw can land in the same bin. 

Consider a subset $T$ of \dmin bins. 
We estimate the probability that there exists a subset of $f{+}1$ throws entirely avoiding $T$. The probability that $\pi_i$ avoids the specific set $T$ entirely is $\binom{\hat{m} - \dmin}{d} \cdot \binom{\hat{m}}{d}^{-1}$~.
 We can bound this probability from above by assuming the selection of $d$ bins is fully independent (i.e., allow repetitions): \begin{align*}
q &=\binom{\hat{m}-\dmin}{d}\binom{\hat{m}}{d}^{-1}\\
&\le \left(\frac{\hat{m}-(2\hat{m}/3+1)}{\hat{m}}\right)^{d} \nonumber < \left(\frac{1}{3}\right)^{d}~.
\end{align*} 


Since the set of $3f{+}1$ throws are independent, this experiment forms a Bernoulli trial with $3f{+}1$ repetitions, each with success probability $q$.

Let $B_f(q) \coloneqq \Pr_{X\sim\operatorname{Bin}(3f{+}1,q)}[X\geq f{+}1]$
denote the binomial upper-tail probability of $f{+}1$ successes. 
Since $q < 1/3$, applying a Chernoff bound yields: \begin{align*}
B_f(q) &\quad\leq \Pr[\operatorname{Bin}(3f{+}1,q) > (3f{+}1)/3] \\
&\quad\leq \exp\left( -(3f{+}1) \cdot D_{\mathrm{KL}}\left(\frac13\middle\|q\right)\right)
\end{align*}
where $D_{KL}$ is KL-divergence with parameters $a = \frac{1}{3}$ and $q$, i.e., $
D_{\mathrm{KL}}\left(\frac13\middle\|q\right)
 = - \frac{1}{3}\ln( \frac{27}{4} q (1 -q)^2)$, 
which leads to: \[
B_f(q) \leq \left[\frac{27}{4} q (1 -q)^2)\right]^{\frac{(3f{+}1)}{3}}~.
\]

The preceding argument applies to a fixed set $T$ of bins. Taking a
union bound over all subsets of \dmin bins bounds the probability that the maximal number $e_{actual}$ of holes left by any subset $S'$ reaches the code distance as follows: 

\[
\Pr[e_{actual} \geq \dmin]\leq
\binom{\hat{m}}{\dmin} B_f(q)
\] 

For $\hat{m}=3m$, $\dmin=2m+1$, Stirling's approximation gives
$\binom{3m}{2m+1} <\binom{3m}{m} \approx (27/4)^m\sqrt{3/(4\pi m)}$~. We use the upper bound $\binom{3m}{2m+1}\leq (27/4)^m$ to obtain: \[
\Pr[e_{actual} \geq \dmin]
\leq
\left(\frac{27}{4}\right)^{m}
\left[\frac{27}{4}
q
\left(1-q\right)^2
\right]^{\frac{(3f{+}1)}{3}}
\] 

The probability above now depends on three parameters: $d$, the number of symbols per throw, $n$, the number of overall parties, and $f$, the fraction of Byzantine ones. This allows to fix $n$ and determine $d$ accordingly; or conversely, fix $d${-}size qShares and determine a required system size $n$; \Cref{sec:concrete}
 derives concrete numbers. 

\paragraph{\textbf{Semi-adaptive adversary}} The bound above assumes a fully adaptive adversary. The semi-adaptive model  forces the adversary to commit to specific $f$ corrupt parties before all parties reveal their $\pi_i$ selections.
For this setting, we are interested in bounding a bad event $e'_{actual}$ occurring within the $2f{+}1$ honest throws only, since the adversary will have already designated $f$ corruptions.  Thus, the bad event $e'_{actual}$ becomes: $f+1$ out of $2f{+}1$ honest throws avoid $\dmin$ bins, and the probability bound on $\Pr[e'_{actual} \geq \dmin]$
can be tightened to:
\[
\Pr[e'_{actual} \geq \dmin]
\leq
\left(\frac{27}{4}\right)^m
\left[
4q
\left(1-q\right)
\right]^{\frac{(2f{+}1)}{2}}.
\]

\paragraph{\textbf{Byzantine–crash adversary}} 
So far, we have considered the case where all $f$ tolerated failures are Byzantine. Practically, we can consider more pragmatic settings and see how we can scale the problem to weaker settings. Thus, we can expand the model to $f=b{+}c$, where $b$ corresponds to Byzantine failures and $c$ to crash failures; $b = f$ corresponds to the usual adversary. Under this model, the adversary can contribute at most $b$ corrupted qShares. A valid qAttest remains consisting of $2f{+}1$ qShares, and it is possible to collect these from parties which do not suffer either type of fault. 

In the fully-adaptive adversary model, the bad event $e''_{actual}$ occurs here when 
$2f{+}1{-}b$ throws out of $3f{+}1$ avoid $\dmin$ bins, resulting in the following probability bound on $\Pr[e''_{actual} \geq \dmin]$: \[
\min\Bigl\{1, \left(\frac{27}{4}\right)^{m} e^{-(3f{+}1)\,D_{\mathrm{KL}}(a\,\|\,q)}\Bigr\}\]
where $a = \frac{2f{+}1-b}{3f{+}1}$. 


A very similar analysis for a semi-adaptive adversary results in the following probability bound: \[
\min\Bigl\{1, \left(\frac{27}{4}\right)^{m} e^{-(2f{+}1)\,D_{\mathrm{KL}}(a'\,\|\,q)}\Bigr\}\]
where $a'  = \frac{2f{+}1-b}{2f{+}1}$.
\section{Concrete Numbers}
\label{sec:concrete}

We evaluate the tradeoff in concrete numbers between the system size $n=3f+1$,
the number $d$ of symbols authenticated per qShare, and the
assignment-failure probability. 
Throughout, we consider
statements of $m=32$ symbols and codewords of length
$\hat{m}=3m$ and minimum code distance $\dmin \coloneqq 2m+1$. Let
$\epsilon_{\mathsf{assign}}$
denote the probability that the assignment leaves at least \dmin holes.
Our default target is
$\epsilon_{\mathsf{assign}}\leq 2^{-128}$.
The reduction in authenticated symbols per party, relative
to authenticating the full statement, is $m/d$.

\subsection{Required system size for a given $d$}
\label{sec:n-min}

We first fix $d$ and determine a sufficient system size
for the default assignment-failure target. We consider
fully adaptive and semi-adaptive adversaries, followed
by mixed Byzantine--crash faults.
Table~\ref{tab:d-vals} summarizes the sufficient thresholds
and the corresponding reductions $m/d$ for all three models.

\paragraph{\textbf{Fully adaptive adversary}}
The bounds in Section~\ref{sec:bounds} are practical when $d\geq 2$.
We evaluate $d\in\{2,3,5,10,21\}$ using the general analysis
for arbitrary $d\geq 2$.

For example, with $d=2$ and $m=32$, the numerical expression
for the bound is approximately
\[
\epsilon_{\mathsf{assign}}
\leq
2^{\,2.754888m-0.251629(3f+1)}.
\]
This gives the sufficient threshold $f\geq 287$.
Increasing $d$ to $3$ substantially reduces the required
system size: the corresponding threshold is $f\geq 103$.

\paragraph{\textbf{Semi-adaptive adversary}}
Under the semi-adaptive model, the analysis also permits
$d=1$. The default target is achieved with $f=1272$
for $d=1$, and with $f=161$ for $d=2$.

\begin{table}[t]
\caption{Concrete parameters and savings under fully
adaptive and semi-adaptive adversary models.
Here $n=3f+1$ and $f=b+c$.}
\label{tab:d-vals}
\centering
\footnotesize
\setlength{\tabcolsep}{1.4pt}
\renewcommand{\arraystretch}{1.25}

\begin{tabular*}{\columnwidth}{
@{\extracolsep{\fill}}lrrrrrrr@{}
}
\toprule
\textbf{Symbols $d$}& \textbf{1}
& \textbf{2} & \textbf{3} 
& \textbf{5} & \textbf{10} & \textbf{21} \\
Reduction $m/d$ & $\times32$ 
& $\times16$ & $\times10.7$ 
& $\times6.4$ & $\times3.2$ & $\times1.5$ \\
\midrule

\rowcolor{gray!24}
\multicolumn{7}{@{}l}{\textbf{Fully adaptive}} \\
\addlinespace[2pt]

\rowcolor{gray!10}
\multicolumn{7}{@{}l}{$b=f$} \\
Parties $n$& --
& 862 & 310  & 127 & 52 & 22 \\
Corrupt $f$ & --
& 287 & 103  & 42 & 17 & 7  \\
\midrule

\rowcolor{gray!10}
\multicolumn{7}{@{}l}{$b=3f/4$} \\
Parties $n$& --
& 493 & 211  & 94 & 40 & 19 \\
Corrupt $f$ & --
& 164 & 70  & 31 & 13 & 6 \\
\midrule

\rowcolor{gray!10}
\multicolumn{7}{@{}l}{$b=f/2$} \\
Parties $n$& --
& 325 & 154  & 73 & 34 & 16 \\
Corrupt $f$ & --
& 108  & 51  & 24 & 11  & 5 \\
\midrule

\rowcolor{gray!24}
\multicolumn{7}{@{}l}{\textbf{Semi-adaptive}} \\
\addlinespace[2pt]

\rowcolor{gray!10}
\multicolumn{7}{@{}l}{$b=f$} \\
Parties $n$& 3817
& 484 & 232 & 109 & 49 & 22 \\
Corrupt $f$ & 1272
& 161 & 77  & 36 & 16 & 7 \\
\midrule

\rowcolor{gray!10}
\multicolumn{7}{@{}l}{$b=3f/4$} \\
Parties $n$ & 1270
& 298 & 160 & 82 & 37 & 16 \\
Corrupt $f$ & 423
& 99 & 53  &27  &12  & 5 \\
\bottomrule

\end{tabular*}
\end{table}

\paragraph{\textbf{Mixed Byzantine--crash faults}}
Keeping the total fault bound $f=b+c$ fixed while reducing
the Byzantine-fault bound $b$ permits more favorable
parameters. As an illustrative case, consider
$b=\lfloor 3f/4\rfloor$.

For the fully adaptive model, the analysis in
Section~\ref{sec:bounds} yields the following
KL-divergence expression, where $q$ is the avoidance
probability defined there:
\[
\epsilon_{\mathsf{assign}}
\leq
\binom{\hat{m}}{\hat{m}/3}
\exp\left(
  -(3f+1)
  D_{\mathrm{KL}}\left(\frac{5}{12}\middle\|q\right)
\right).
\]
Using $\hat{m}=3m$, this becomes
\[
\epsilon_{\mathsf{assign}}
\leq
\left(\frac{27}{4}\right)^m
\left[
  \frac{12^{12}}{5^5 7^7}
  q^5(1-q)^7
\right]^{\frac{(3f+1)}{12}}.
\]
The corresponding expression for the semi-adaptive model is 
\[
\epsilon_{\mathsf{assign}}
\leq
\left(\frac{27}{4}\right)^m
\left[
  \frac{8^8}{5^5 3^3}
  q^5(1-q)^3
\right]^{\frac{(2f+1)}{8}}.
\]
\subsection{Varying the assignment-failure target}

A uniform assignment-failure target of $2^{-128}$ may be unnecessarily conservative. Applications could instead require this bound in the common case while allowing graceful degradation as the number of Byzantine faults increases, tolerating a higher assignment-failure probability in extreme but rare fault configurations. This flexibility can enable smaller qShares for a fixed system size and fault bound, or smaller systems for a fixed $d$.

\begin{figure}[!htbp]
    \centering

    \subfloat[Fully-adaptive adversary.\label{fig:fully-adaptive}]{
        \includegraphics[
            width=\linewidth
        ]{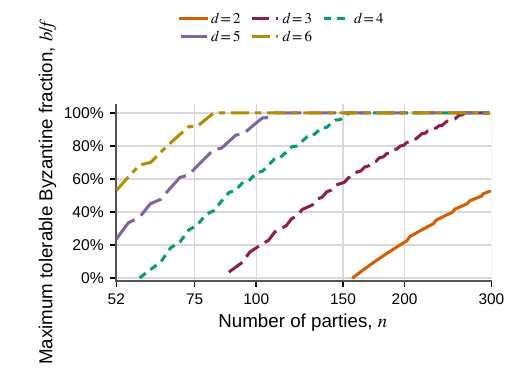}
    }

    \vspace{-1mm}

    \subfloat[Semi-adaptive adversary.\label{fig:semi-adaptive}]{
        \includegraphics[
            width=\linewidth
        ]{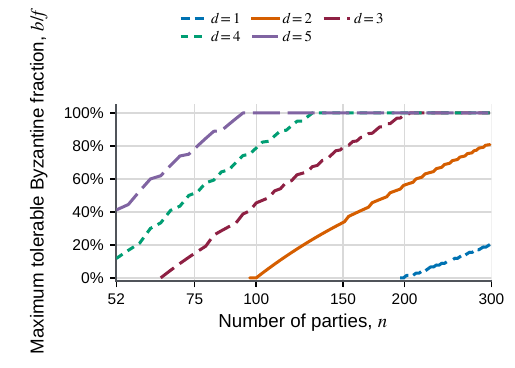}
    }

    \vspace{1mm}

    \caption{
Maximum tolerable Byzantine fraction $b/f$ as a function
of the number of parties $n$, under (a) fully adaptive and
(b) semi-adaptive adversaries, for varying $d$, with $n=3f+1$
and $m=32$.
For each configuration, the maximum Byzantine tolerance is
determined using the refined binomial-tail union bound with a
security target of $2^{-128}$. Curves connect discrete $n=3f+1$
configurations by straight segments as visual guides.
Values of $b/f=100\%$ indicate that the bound is satisfied
for the full Byzantine fault budget $b=f$.
    }
    \label{fig:forgeability-varying-d}
\end{figure}

\begin{figure}[t]
    \centering
    \includegraphics[width=\columnwidth]{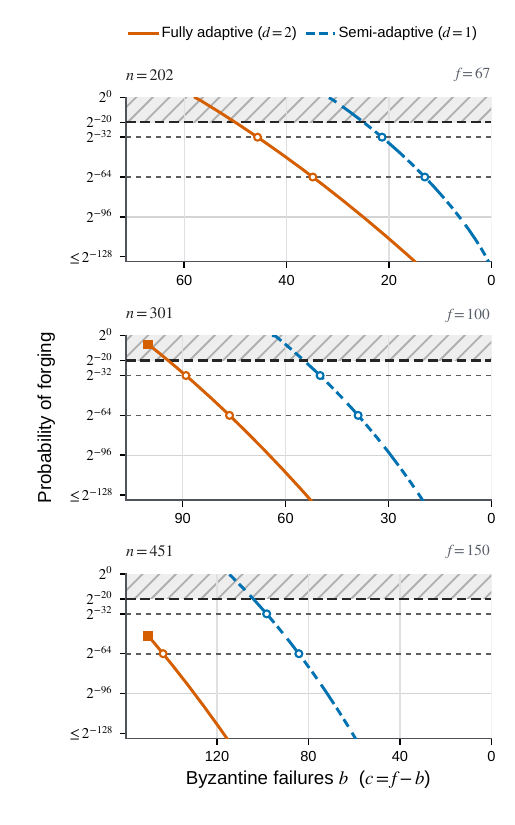}
    \vspace{-6mm}
    \caption{Probability of forging as a function of Byzantine failures considering a large-scale number of parties $n$ under fully adaptive and semi-adaptive adversaries for smallest corresponding $d$, with $m=32$. Curves terminate when the bound reaches 1 or $2^{-128}$.}
    \label{fig:forgeability}
\end{figure}

Figure~\ref{fig:forgeability-varying-d} shows how the proportion of possible Byzantine failures for target security
varies with $d$ for fixed system sizes.
Figure~\ref{fig:forgeability} shows the effect of varying
the Byzantine-fault bound.

\captionsetup[subfloat]{font=footnotesize}

\subsection{Selecting parameters}

The general bounds allow parameter
selection beyond the examples above. Given an adversary
model, system size, fault bounds, and assignment-failure
target, we select the smallest $d$ satisfying the
corresponding bound.

For small systems, e.g., around $n \leq 70$, the simple deterministic construction
from the introduction provides a useful point of comparison:
each party authenticates approximately two-thirds of the
statement, giving an asymptotic reduction of $1.5\times$.
The randomized construction offers progressively larger
reductions as the system grows, exceeding $10\times$ in
suitable parameter regimes.

Parameter selection should
therefore compare the randomized bounds with the
deterministic construction and use whichever authenticates
fewer symbols per party.

\section{Related Works} 

Since \QUASI lies at the intersection of (1) threshold and aggregate signatures, (2) hash-based one-time signatures, and (3) quorum protocols, we review related work in these three dimensions.

\textbf{Threshold signatures.} Desmedt et al.~\cite{DesmedtFrankel91} introduced the concepts of \textit{threshold signatures} which relies on parties holding shares of a single signing key associated with a common public key. The key idea is that each party produces a partial signature, where any collection of $t$ valid shares can be combined to create a standard signature, without revealing the signers. The cost of threshold signature schemes is that they necessitate pre-coordination and sharing of the signing key among parties.
Examples of threshold signature schemes include threshold RSA~\cite{Shoup2000}, threshold BLS~\cite{Boldyreva2003}, threshold ECDSA~\cite{GennaroGoldfeder2020}, and FROST (threshold Schnorr)~\cite{KomloGoldberg2020}. 
Micali et al. extended this to the notion of \textit{multisignatures} in~\cite{MicaliOhtaReyzin01} where each party holds an independent signing key and uses it to contribute a signature. These signatures are then collected into a multisignature certifying that a specified set of parties signed the same message. Verification generally requires public keys for specific participants. Hence signers are known and no shared secret key or threshold secret-sharing structure is used. 
Examples of multisignature schemes include Bellare--Neven (BN)~\cite{BellareNeven2006}, MuSig~\cite{MaxwellPoelstraSeurinWuille2019}, MuSig2~\cite{NickRuffingSeurin2021}, and BLS multisignatures~\cite{BonehDrijversNeven2018}.
Extending to signatures on potentially different messages, Boneh et al.\ introduced \textit{aggregate signatures} in~\cite{BonehGentryLynnShacham03}. When all parties sign the same message, aggregate signatures closely resemble multisignatures.  
Examples of aggregate signature schemes include BGLS~\cite{BonehGentryLynnShacham03}, LMRS sequential aggregate signatures~\cite{LysyanskayaMicaliReyzinShacham04}, and LOSSW sequential aggregate signatures~\cite{LuOstrovskySahaiShachamWaters06}.

While not being a threshold-signature scheme, QUASI is a newly conceived hybrid primitive resting on intersection of the above mentioned schemes: it supports quorum-based aggregation, partial signatures are bound to their respective signers, and it allows any threshold of valid, non-conflicting shares satisfying the quorum requirement to create a finalized attestation. 

More recent work extends threshold signing to post-quantum hash-based signatures.
Kelsey, Lang, and Lucks~\cite{kelsey2025threshold} present an early
construction in which a trusted dealer generates all one-time
signing secrets in advance, distributes shares, and publishes
auxiliary data. An aggregator then combines the parties'
contributions to reconstruct the signature values selected by
the message.
PRAWNS~\cite{BonehEtAl2026PRAWNS} follows a related
preprocessing-and-reconstruction approach based on threshold
pseudorandom functions (PRFs). Its lattice-based instantiation
supports large groups, while its purely hash-based instantiation
targets smaller committees.

QUASI takes a fundamentally different and simpler approach than both: independently
generated keys and sparse symbol authentication enable
dealer-free attestation for large groups of parties.
QUASI's weak-$f$-unforgeability shares with PRAWNS's
context-aware construction and Haystack's message-binding
mechanism the goal of preventing malicious recombination
of contributions into an unauthorized result.

\textbf{Hash-based Signatures.}
The notion of one-time signatures (OTS) was first introduced by Lamport in~\cite{LOTS}. The key idea is that signatures are constructed by
revealing secret values whose hashes form the public key.
Merkle subsequently developed hash-based signature
constructions\cite{WOTS}, allowing multiple one-time
public keys to be authenticated under a single Merkle-tree
root. Further improvements of Winternitz one-time signatures (WOTS) and their
extension WOTS+~\cite{hulsing2013wotsplus} use hash chains to reduce signature size
at the expense of additional hash computations. These constructions provide a foundation for post-quantum signatures based on cryptographic hash functions, where examples include
XMSS~\cite{xmss} and LMS/HSS~\cite{RFC8554}.
Both are stateful schemes that require tracking used
one-time keys to prevent reuse, and are standardized
in NIST SP~800-208~\cite{SP800-208}.
In contrast, SPHINCS+, standardized as SLH-DSA
in FIPS~205~\cite{slhdsa}, provides stateless
hash-based signatures by combining FORS, WOTS+,
and a hypertree, eliminating explicit signing-state
management at the cost of larger signatures. More recent advancements include leanSig and leanXMSS have been proposed
for post-quantum blockchain consensus~\cite{LeanSigNote,EIP8450},
with leanVM~\cite{leanVM} supporting proof-based aggregation
of signature verification. However, none of these hash-based signatures inherently support compact aggregation
of signatures produced by independent signers.

A significant body of work uses hash-based zero-knowledge proofs,
such as STARKs~\cite{BenSassonEtAl18}, to prove the
correctness of signature verification without revealing
the underlying signature or public key.
By committing to an execution trace using Merkle trees
and verifying algebraic constraints, these systems
can certify the validity of classical signatures,
such as ECDSA~\cite{RiscZero}, while keeping their public keys hidden.
This approach has been explored for post-quantum
migration~\cite{PQMigration,RiscZero}. More recent advances in succinct proof systems, including PLONK~\cite{plonk}, its hash-based extensions such as Plonky2~\cite{Plonky2}, and Flock~\cite{flock}, have substantially improved proving efficiency. In particular, Flock demonstrates efficient batch proving of cryptographic hash computations. 

\QUASI
targets shares of parties that can be combined in the compact attestation, without requiring a shared threshold signing key or general-purpose zero-knowledge proof. However, \QUASI is orthogonal to, and compatible with, proof-based compression. Putting the two together compounds the compression benefits of both.

\textbf{Quorum-based protocols.}
Byzantine quorum systems~\cite{MalkhiReiter98} provide a fundamental building block for fault-tolerant distributed systems. In the classical setting of $n=3f+1$ parties, quorums of size $2f+1$ intersect in at least $f+1$ parties, guaranteeing at least one honest party in their intersection. Beyond consensus, quorum-based constructions underpin Byzantine reliable broadcast~\cite{bracha1987}, fault-tolerant storage and register emulation~\cite{MalkhiReiter98}, and payment systems such as FastPay~\cite{fastpay}, Stingray~\cite{stingray}, and the Sui blockchain~\cite{sui-lutris}.

Quorum intersection is also central to Byzantine fault-tolerant consensus. The pioneering Practical Byzantine Fault Tolerance (PBFT) protocol of Castro and Liskov~\cite{pbft} established a practical quorum-based approach to Byzantine consensus. Numerous subsequent protocols have built upon this foundation, and a comprehensive survey is beyond our scope. Notably, HotStuff~\cite{hotstuff} uses threshold signatures to aggregate quorum votes into compact quorum certificates (QCs), enabling linear communication complexity in the optimistic case. It remains an open question whether sub-linear aggregation of hash-based signatures is possible in HotStuff-like protocols.

More recently, information-theoretic BFT protocols, including IT-HotStuff~\cite{IT-hotsuff}, TetraBFT~\cite{tetrabft}, Forget-IT~\cite{forgetit}, Simple-IT~\cite{simpleit}, 
Quintus~\cite{quintus}, Quantum Cadence~\cite{quantumcadence}, and others, 
solved consensus without digital signatures, relying instead on authenticated point-to-point communication. While these protocols avoid public-key signature operations during consensus, they do not inherently produce transferable, publicly verifiable certificates that external clients can independently verify.

These developments highlight the importance of efficient external attestation for quorum-based systems. \QUASI fills this role by enabling parties to produce publicly verifiable post-quantum attestations of statements already validated by a quorum, without modifying the underlying consensus protocol.
Thus, \QUASI provides a complementary mechanism for externally certifying the outcomes of both traditional and signature-free quorum protocols.

\section{Discussion}

QUASI opens many directions for further research, from tighter
analysis and improved constructions to practical optimizations
and new applications.

The construction and analysis of~\Cref{sec:bounds} leave substantial room to improve
the constants. In particular, the union bound used to link $d$ and $f$
is fairly crude; a tighter analysis may yield smaller qShares
or support smaller systems. Applications may also tolerate an
assignment-failure probability higher than $2^{-128}$, allowing
more favorable parameters. Likewise, corruption thresholds
other than $n/3$ merit exploration.

We explored a randomized construction that expands the statement
into a codeword to tolerate flipped symbols. We are confident
that deterministic constructions which tolerate symbol flips can also achieve significant
benefits beyond the simple assignment of each symbol to $2f+1$
parties, particularly for medium-size systems. We are currently
exploring improved deterministic symbol assignments.

A practical implementation would provide end-to-end measurements
and may reveal further opportunities for improvement by leveraging
the underlying QUASI one-time hash-based authentication and
its reduced size. These opportunities include optimizing
stateful and stateless key management and amortizing Winternitz
checksum symbols across qShares. The smaller per-party
contributions also motivate tuning and optimizing SNARK-based
aggregation, with end-to-end comparisons against proving the
verification of collections of full hash-based signatures.

QUASI may also be useful within conventional signature-based
consensus protocols, beyond serving as an external attestation
layer. The reason is that hash-based authentication is simple and computationally
efficient compared with conventional digital signatures, while their main hurdle, size, is removed by QUASI.  Such integration would
require examining which protocol steps can use QUASI's weaker
guarantees while preserving the consensus security.

Finally, QUASI introduces a new foundation for group-based
attestation. Its lower bounds, security properties, and
relationship to threshold signatures, aggregate signatures,
and multisignatures merit further study. Exploring these
connections may reveal applications in other cryptographic
settings and clarify the tradeoffs between attestation
guarantees and efficiency.
\section{Acknowledgment }
We thank Benjamin Marsh and Nibesh Shrestha for their valuable feedback and suggestions that helped improve this paper.

\clearpage
\newpage
\bibliographystyle{acm}
\bibliography{references}

\clearpage
\newpage
\appendices

\section{} \label{app:qots}
To authenticate an ordered subset of $d$ codeword symbols, QUASI concatenates them into a word and applies standard WOTS signing and verification. We refer to this adaptation as \emph{Quasi One-Time Authentication Scheme} (QOTS), which inherits the one-time unforgeability guarantees of WOTS.

\paragraph{\textbf{Quasi One-Time Authentication Scheme (QOTS)}}
\label{con:qots}
Let $\mathcal{A}=\{0,\ldots,2^\omega-1\}$ be the symbol alphabet. Let $a = \left(a_1, \dots, a_d \right) \in \mathcal{A}^d$ denote the ordered word authenticated by QOTS. In the outer QUASI construction, party $p_i$ obtains this word by selecting positions from the codeword $c(M)$ according to the mapping $\pi_i:[d]\to[\hat m]$:
\[
 a := c(M)_{\pi_i(1)}\cdots c(M)_{\pi_i(d)}\in\mathcal{A}^{d}.
\]
Let
$\Pi_{\mathsf{WOTS}}
=(\textsf{Setup}_{\mathsf{WOTS}},\textsf{KeyGen}_{\mathsf{WOTS}},
\textsf{Sign}_{\mathsf{WOTS}}, \\
\textsf{Verify}_{\mathsf{WOTS}})$
be a one-time signature scheme for words in $\mathcal{A}^{d}$.

\paragraph{\textbf{WOTS encoding}}
For a word $a = \left(a_1, \dots, a_d \right)$, WOTS appends the fixed-length
base-$2^\omega$ encoding of the checksum
$C(a)=\sum_{j=1}^{d}(2^\omega-1-a_j)$.
The checksum requires
$t=\left\lfloor \frac{\log_2\left(d(2^\omega-1)\right)}{\omega} \right\rfloor+1$ symbols.
We can then write the resulting encoding as 
$E(a)= (a_1, \dots , a_d, s_1, \dots s_t) =(e_1,\ldots,e_{d+t})$, where  $ s_1, \dots, s_t$ are base-$2^\omega$ checksum digits. Following the basic WOTS construction, the scheme samples an independent secret starting value $x_i$  
for each encoded symbol. The scheme constructs hash chains of length $2^\omega$, with each public endpoint computed as $y_i = H^{2^\omega-1}(x_i)$. The signature will be defined as $\tau_i=H^{e_i}(x_i)$. To verify the signature, it is sufficient to check: \[
  \forall i\in[d+t]:
  \qquad H^{2^\omega-1-e_i}(\tau_i)=y_i.
\]
Here, $H^r$ denotes $r$ successive hash applications of the hash function $H$. Alternatively, a concrete instantiation may instead use the addressed, randomized chains of WOTS+, with its corresponding security assumptions and verification algorithm.

\paragraph{\textbf{QOTS interface}}
The scheme
$\Pi_{\mathsf{QOTS}}=(\textsf{Setup}_{\mathsf{QOTS}},\textsf{KeyGen}_{\mathsf{QOTS}},\textsf{Sign}_{\mathsf{QOTS}},\textsf{Verify}_{\mathsf{QOTS}})$
is defined as follows. In the terminology of \QUASI{}, its authentication output is a
\emph{qShare}.

\begin{itemize}
\item $\textsf{Setup}_{\mathsf{QOTS}}(1^\kappa,d)\to\mathsf{pp_{\mathsf {QOTS}}}$.
Run $\mathsf{pp}_{\mathsf {WOTS}}
\leftarrow\textsf{Setup}_{\mathsf {WOTS}}(1^\kappa)$
for words in $\mathcal{A}^{d}$.
Output
$\mathsf{pp_{\mathsf {QOTS}}}=(\mathsf{pp}_{\mathsf {WOTS}},2^\omega,d)$.

\item $\textsf{KeyGen}_{\mathsf{QOTS}}(\mathsf{pp_{\mathsf {QOTS}}})
       \to(\mathsf{vk^{QOTS}},\mathsf{sk^{QOTS}})$.
Generate
$(\mathsf{vk}_{\mathsf {WOTS}},\mathsf{sk}_{\mathsf {WOTS}})
\leftarrow\textsf{KeyGen}_{\mathsf {WOTS}}(\mathsf{pp}_{\mathsf {WOTS}})$.
Output
$\mathsf{vk^{QOTS}}=\mathsf{vk}_{\mathsf {WOTS}}$ and
$  \mathsf{sk^{QOTS}}= \mathsf{sk}_{\mathsf {WOTS}}$.

\item $\textsf{Sign}_{\mathsf{QOTS}}(\mathsf{pp_{QOTS}},\mathsf{sk^{QOTS}},a)\to\tau$.
Check that $a\in\mathcal{A}^d$ and output
\[
  \tau\leftarrow
  \textsf{Sign}_{\mathsf {WOTS}}
  (\mathsf{pp}_{\mathsf {WOTS}},\mathsf{sk}_{\mathsf {WOTS}},a).
\]

\item $\textsf{Verify}_{\mathsf{QOTS}}(\mathsf{pp_{QOTS}},\mathsf{vk^{QOTS}},a,\tau)
       \to\{0,1\}$.
Reject malformed inputs. Otherwise, output
\[
  \textsf{Verify}_{\mathsf {WOTS}}
  (\mathsf{pp}_{\mathsf {WOTS}},\mathsf{vk}_{\mathsf {WOTS}},a,\tau).
\]
\end{itemize}

\paragraph{\textbf{Correctness}}
For every valid $a\in\mathcal{A}^d$ and every
honestly generated key pair, an honestly generated QOTS signature $\tau$
verifies with probability $1$.
Indeed, signing and verification apply WOTS to the same word
$a$.
For the basic hash-chain construction, each component satisfies
\[
  H^{2^\omega-1-e_i}\bigl(H^{e_i}(x_i)\bigr)
  = H^{2^\omega-1}(x_i)=y_i.
\]
Thus, correctness follows directly from WOTS correctness, as completing the hash chain from each signature component yields its corresponding public endpoint. 

\paragraph{\textbf{One-time unforgeability on selected symbols}} An adversary receives $\mathsf{vk^{QOTS}}$ and may request a qShare
on one chosen word $a$.
It wins by producing $(a^*,\tau^*)$ such that it passes verification
$\textsf{Verify}_\mathsf{{QOTS}}(\mathsf{pp_{QOTS}},\mathsf{vk^{QOTS}},a^*,\tau^*)=1$
and $a^*\neq a$.
QOTS is one-time unforgeable on selected symbols if every
probabilistic polynomial-time adversary wins with probability
negligible in $\kappa$.

\paragraph{\textbf{Security argument}}
Any successful QOTS forgery immediately yields a WOTS forgery.
A reduction embeds its WOTS verification key as 
$\mathsf{vk^{QOTS}}=\mathsf{vk_{\mathsf WOTS}}$, answers the sole
QOTS signing query by requesting a WOTS signature on $a$,
and outputs $(a^*,\tau^*)$.
This is a valid signature on a different word since $a^* \neq a$.
Consequently, QOTS inherits one-time unforgeability from WOTS,
with the same success probability.

The role of the checksum parallels that of the unrevealed
preimages in Lamport-OTS.
An adversary can advance a revealed hash-chain value, thereby
increasing its encoded symbol.
However, any distinct word either decreases a data symbol or,
if all data symbols only increase, decreases the checksum.
In either case, some encoded symbol decreases, requiring a
chain value that cannot be obtained by merely advancing the
revealed values.
The underlying WOTS security theorem formalizes this argument
under its hash-function assumptions.

\paragraph{\textbf{Scope and one-time restriction}}
A qShare authenticates only $a \in \mathcal{A}^d$. In the outer QUASI construction, this word consists of the codeword symbols selected by mapping $\pi_i$. Certification of the complete statement is provided by the
quorum construction.
Each QOTS key pair must be used for at most one signing query.
Post-quantum security follows from the same reduction when
the underlying WOTS instantiation is secure against
quantum polynomial-time adversaries.

\end{document}